# An Inter-Component Pixels Permutation Based Color Image Encryption Using Hyper-chaos

**Musheer Ahmad[1] and Hamed D Al-Sharari[2]**
[1]Department of Computer Engineering, Faculty of Engineering & Technology, Jamia Millia Islamia, New Delhi, India
E-mail: musheer.cse@gmail.com
[2]Department of Electrical Engineering, College of Engineering, AlJouf University, AlJouf, Kingdom of Saudia Arabia
E-mail: hamed_100@hotmail.com
Tel: +91-112-698-0281; Fax: +91-112-698-1261

**Abstract**

In this paper, a simple and robust color image encryption algorithm based on high-dimensional chaotic maps is proposed. The algorithm employs a 3D Arnold transform to perform inter-component shuffling of plain-image, while a 2D hyper-chaotic map is used to confuse the relationship between the encrypted image and plain-image. The control parameters of Arnold transform are extracted from pending plain-image, thereby establishing a dependency to the plain-image. The confusion process is carried out in cipher-block chaining mode to make it dependent to the encrypted image. This makes the algorithm able to resist the cryptographic attacks. The performance of the proposed algorithm is analyzed through computer simulations. The experimental analyses show that the proposed algorithm has desirable properties of high security, robustness to cryptographic attacks and practicability to protect multimedia color images.

## 1. Introduction

Chaos based image encryption is one of recently and widely used approaches to fulfill the security requirements of digital images. The chaotic systems are very sensitive to initial conditions and system parameters. For a given set of parameters in chaotic regime, two close initial conditions lead the system into divergent trajectories. Therefore, an image encryption and decryption scheme can be developed if the initial condition and parameters are chosen as keys [1]. Since the same parameters are used for encryption and decryption, the chaos based image encryption scheme is symmetric. Chaos based image encryption schemes are considered good for practical use because they have important characteristics like (i) they are very sensitive to initial conditions and system parameters, (ii) they have pseudo-random property and non-periodicity as the chaotic signals are usually noise-like, etc. All these characteristics make chaos-based encryption algorithm an excellent and robust cryptosystem against any statistical attacks [2]. A number of image encryption algorithms based on chaotic maps have been proposed in recent past. For image encryption, two-dimensional or higher-dimensional chaotic systems are employed as the image can be considered as a 2D array of pixels [3-16].

In this paper, a new color image encryption algorithm is proposed. Three-dimensional Arnold map is used to shuffle the positions of pixels. A hyper-chaotic map is employed to confuse the relationship between the encrypted image and plain-image. The shuffling process is made plain-image dependent and confusion process through hyper-chaotic map is made encrypted image dependent, by executing it in the cipher-block-chaining mode of encryption. This makes the algorithm robust against cryptographic potential chosen-plaintext and known-plaintext attacks. Rest of this paper is organized as follows: the proposed algorithm is described in Section 2. The simulation analyses and results are reported in Section 3, followed by the conclusion of the work drawn in Section 4.

## 2. Proposed Algorithm

The proposed algorithm explored the inherit advantages of the 3D Arnold transform and 2D hyper-chaotic map. A 3D Arnold transform is an extended Arnold map with four control parameters [6] is as follows:

$$\begin{pmatrix} x' \\ y' \\ z' \end{pmatrix} = \begin{pmatrix} 1 & a & 0 \\ b & ab+1 & 0 \\ c & d & 1 \end{pmatrix} * \begin{pmatrix} x \\ y \\ z \end{pmatrix} \mod \begin{pmatrix} M \\ N \\ 3 \end{pmatrix} \quad (1)$$

Where a, b, c and d are four control parameters which are positive integers and (x', y', z') is the new position of the original pixel position (x, y, z) of color plain-image when Arnold transform is applied once to the plain-image. A 3D Arnold transform apparently permutes/shuffles the organization of pixels of color plain-image by replacing the position of the image pixel points with new coordinate. The scrambling with Eqn (1) results into an inter-component pixels scrambling. Consequently, the pixels of three red, green and blue components of the image are inter-mixed, resulting in better perceptual degradation of the scrambled image which further complicates the works of cryptanalysts. Now, after the application of several iterations (*n*), the correlation among the adjacent pixels is disturbed completely and the image appears distorted and meaningless. But, after iterating many times it will return the plain-image i.e. the Arnold transform is periodic. The periodicity of Arnold transform degrades the security of the encryption, as the possible attacks may iterate the Arnold transform continuously to re-appear the original plain-image. To deal with the periodicity attack, a 2D hyper-chaotic map is employed to change the gray values of the scrambled image so as to further improve the security of algorithm.

In chaotic system, there is only one positive Lyapunov exponent. Messages masked through such simple systems are not always have the sufficient security. It is suggested that this problem can be overcome by using hyper-chaotic systems. Hyper-chaotic systems have better and more complex dynamical features as compared to normal chaotic systems. They have more than one positive Lyapunov exponents unlike chaotic systems. As a result, hyper-chaos based approach boosts the security capabilities of security systems. Due to its higher randomness and unpredictability than chaotic system, the hyper-chaotic systems are more useful and credential for chaos-based cryptography [10]. A two-dimensional hyper-chaotic system with high complexity [17] is described below in Eqn (2).

$$\begin{aligned} x_{n+1} &= a_1 x_n - a_2 y_n^2 \\ y_{n+1} &= a_3 x_n - a_4 y_n \end{aligned} \quad (2)$$

Where, $a_1$, $a_2$, $a_3$ and $a_4$ are system parameters. The system has two positive lyapunov exponents, so it is a hyper-chaotic. The statistical analysis of the generated *x* and *y* sequences shows that they have poor balance, autocorrelation and cross-correlation properties. The mean values are

*mean(x)*= -0.1697 and *mean(y)*= 0.2074. To improve the statistical properties of sequences generated by Eqn(2), the preprocessing described in Eqn(3)-(4) is performed. The range of preprocessed values are $0 < x, y < 1$ and the averages after preprocessing are *mean(x)* = 0.4994 and *mean(y)* = 0.4974, which are closer to the ideal value 0.5. Now, the preprocessed sequences have better balance distribution, correlation properties and they can be used in to build a better and strong cryptographic system.

$$x_i = 10^6 x_i - floor(10^6 x_i) \quad (3)$$

$$y_i = 10^6 y_i - floor(10^6 y_i) \quad (4)$$

The whole encryption process consists of following steps of operations.

**Step 1.** Evaluate the sum of all the gray-values of color plain-image $P$ of size M×N×3, let it be *gvSum*.

**Step 2.** Extract the control parameters of Arnold transform from *gvSum* as:
  $a = 13 + (gvSum)\bmod(97)$
  $b = 23 + (gvSum)\bmod(59)$
  $c = 17 + (gvSum)\bmod(79)$
  $d = 37 + (gvSum)\bmod(43)$
  $n = 07 + (gvSum)\bmod(31)$

**Step 3.** Shuffle the plain-image $P$ using 3D Arnold transform with parameters $a$, $b$, $c$, $d$ and $n$.

**Step 4.** Let the shuffled color image obtained is $S$. Reshape $S$ into 1D sequence

**Step 5.** Iterate the hyper-chaotic system given in Eqn(2), with parameters $x_0$, $y_0$, $a_1$, $a_2$, $a_3$ and $a_4$, for M×N×3 times to get $X$ and $Y$ sequence. For each iteration $i$, we get values $x_i$ and $y_i$

**Step 6.** Preprocess the two sequences using Eqn(3)-(4) to improve their statistical properties.

**Step 7.** The preprocessed $x_i$ and $y_i$ are multiplied by $10^{14}$, extract their integer parts, then apply modulo-256 operation to get values in the range of 0 – 255 corresponding to each $x_i$ and $y_i$. Let the converted sequence be $K_1$ and $K_2$ having size of M×N×3. Choose $C(0)$ in the range of 0 – 255.

**Step 8.** Applying XOR operation in following way using CBC mode to encrypt all pixels.
  $C(i) = S(i) \oplus K_1(i) \oplus K_2(i) \oplus C(i-1)$

Where $i = 1, 2, \ldots, M\times N\times 3$, Reshape 1D sequence $C(i)$ to a color encrypted image. The decryption progresses similar to the encryption one, described above, but in the reverse order.

## 3. Experimental Results

In this section, experimental analysis of the proposed algorithm has been done. The same standard *Lena* color image of size 256x256×3 is opted to make performance comparisons with existing color image encryption algorithms. The original image and its histograms are shown in Figure 1. The initial conditions of 2D hyper-chaotic map taken for encryption are: $x_0 = 0.2159$, $y_0 = 0.5738$, $a_1 = 1.55$, $a_2 = 1.3$, $a_3 = 1.1$ and $a_4 = 0.1$. The initial conditions of hyper-chaotic map constitute the secret keys of proposed algorithm. The color image obtained after encrypting the scrambled image is shown in Figure 2(a). The gray value distributions of its R, G, B components are shown in Figure 2(b)-(d). From Figure 2, we can see that the histogram of the encrypted image are fairly uniform and has a random-like appearance. Hence, the gray values distributions of encrypted image are uniform and flat like a noise-image.

**Figure 1.** Standard plain-image *Lena* and histograms of its R, G, B components

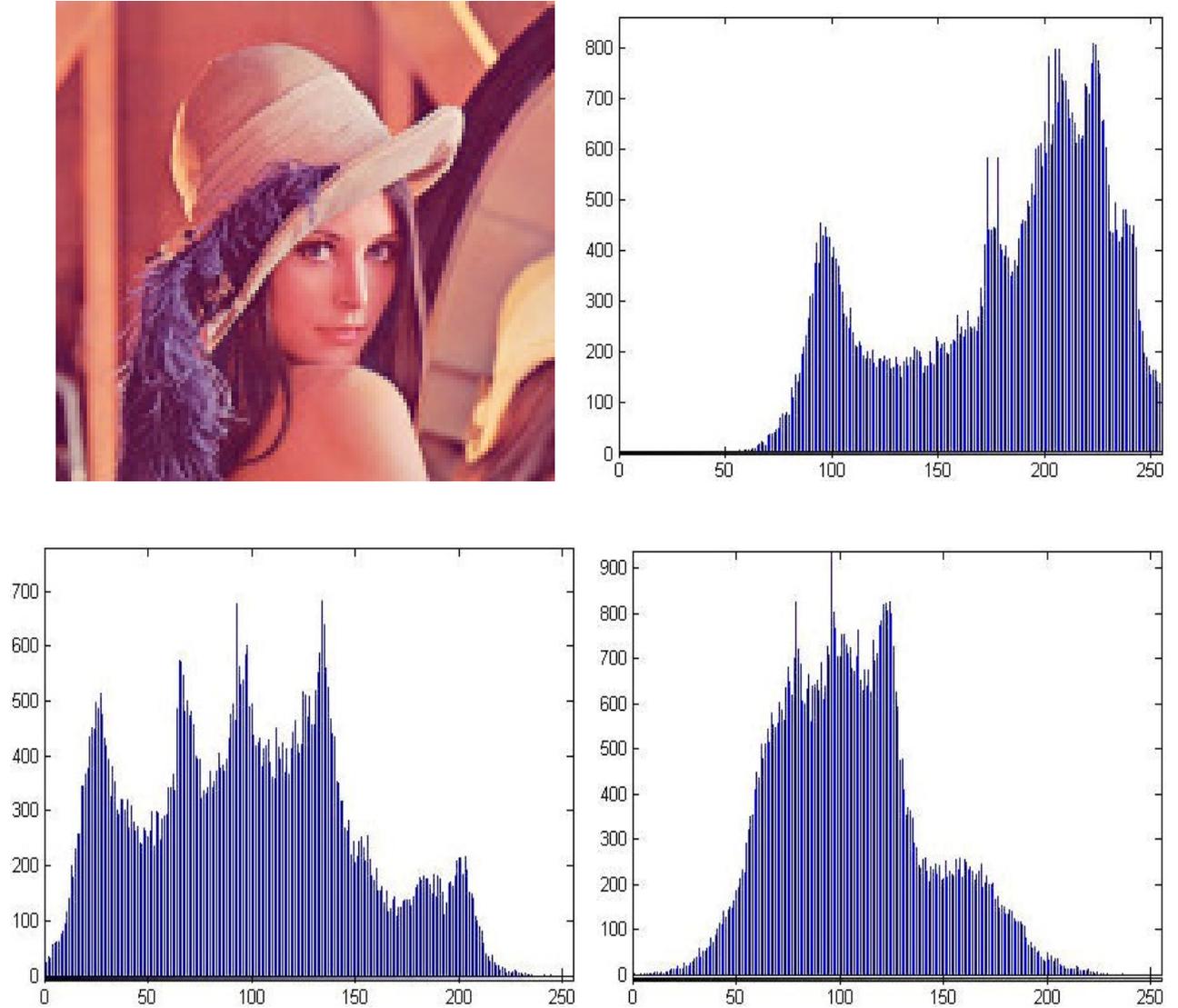

## 3.1. Chi-square Test

The performance of a security system can be estimated through Chi-square test [13]. It is a statistical test used to examine the variations of data from the expected value. It is defined as:

$$\chi^2 = \sum_{i=1}^{256} \frac{(O_i - E_i)^2}{E_i} \quad (4)$$

Where *i* is the number of gray values, *Oi* and *Ei* are observed and expected occurrence of each gray value (0 to *255*), respectively. The less the value of chi-square $\chi^2$ better will the encryption performance of algorithm. The values of $\chi^2$ for R, G, B components of plain-image and encrypted images are depicted in Table 1. We get extremely low values of chi-square as compared to the original image and other encrypted images. These lower values of $\chi^2$ depict that the proposed algorithm provides fairly good encryption effect.

**Figure 2.** Encrypted *Lena* image and histograms of its R, G, B components

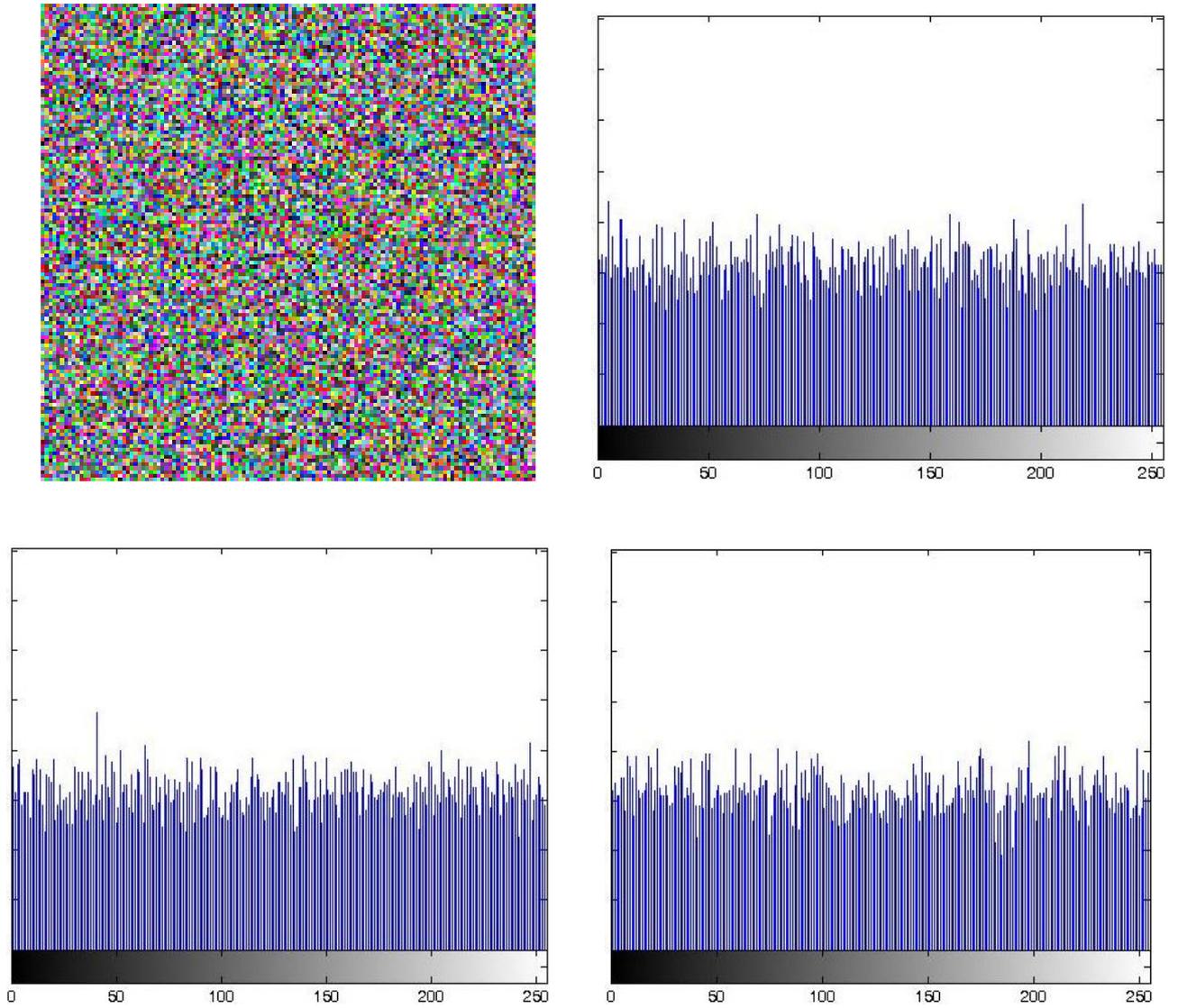

**Table 1.** Chi-square values for original and encrypted images

| Method | R | G | B |
| --- | --- | --- | --- |
| Original | 65274.12 | 30609.43 | 91931.33 |
| Rhouma *et al.* [14] | NA | NA | NA |
| Hongjun *et al.* [15] | NA | NA | NA |
| Huang *et al.* [16] | 12343.32 | 8570.18 | 37977.49 |
| Proposed | 309.73 | 268.86 | 283.54 |

## 3.2. Pixels Correlation Test

The digital multimedia data has high correlations among their adjacent pixels/frames. An encryption algorithm should be able to eliminate the correlations among the vertically, horizontally and diagonally adjacent pixels in images. Highly correlated pixels have a value of coefficient as ±1, where as, the uncorrelated pixels have correlation coefficient close to 0. Pixels correlations in original and other encrypted images for three directions are provided in Table 2. The correlation scores show that the adjacent pixels in encrypted images are closer to 0, this means that the proposed algorithm perfectly de-correlates the adjacent pixels in encrypted image in better way and fulfills the requirement of an efficient encryption system.

Table 2. Correlation coefficient for original and encrypted images

| Method | Horizontal | Vertical | Diagonal |
|---|---|---|---|
| Original | 0.9597 | 0.9792 | 0.9570 |
| Rhouma et al. [14] | 0.0845 | 0.0681 | NA |
| Hongjun et al. [15] | 0.0965 | -0.0318 | 0.0362 |
| Huang et al. [16] | 0.1257 | 0.0581 | 0.0226 |
| Proposed | 0.0076 | 0.0083 | 0.0059 |

## 3.3. Key Sensitivity Test

In our experimental analysis the precision is $10^{-14}$ and algorithm is sensitive to a tiny change in the secret keys. If we change a little ($10^{-14}$) any of the initial conditions then the encrypted image is totally different from the original image. The simulation analysis is carried to calculate the percentage difference between the image shown in Figure 2(a) and the encrypted image(s) obtained with tiny change in one of the secret key parameter. It has been experimentally verified that, in all cases, more than 99% difference is achieved. Hence, the proposed algorithm is highly sensitive to a small change in secret key.

## 3.4. Entropy Test

The information entropy [18] gives a measure of statistical randomness and distortion taken place in encrypted images. It is highly desirable from security viewpoint that encrypted images should have their entropy measures close to the ideal value. The entropy measures for the R, G, B components of original and encrypted images are shown in Table 3. It is evident from the Table that the values for proposed algorithm are better as compared to the existing color image encryption algorithms, as the values are more close to the ideal value 8. This implies that the information leakage in the encrypted images is negligible and encrypted image is secure against entropy-based attack.

## 3.5. Net Pixels Change Rate Test

Net pixels change rate (NPCR) is a parameter by which the encryption effect can be estimated. It is described by the following equation.

$$N(P,C) = \frac{\sum_{i,j} D(i,j)}{M \times N} \times 100\% \quad \text{where} \quad D(i,j) = \begin{cases} 0 & C(i,j) = P(i,j) \\ 1 & C(i,j) \neq P(i,j) \end{cases}$$

The *P(i,j)* and *C(i,j)* are the plain-image and corresponding encrypted image, respectively. The NPCR obtained for three images under test are 99.49%, 99.56% and 99.62% for R, G, B components, respectively. The quantified scores of NPCR illustrate that the proposed algorithm offers high encryption effect.

Table 3. Information entropy for original and encrypted images

| Method | R | G | B |
| --- | --- | --- | --- |
| Original | 7.2359 | 7.5689 | 6.9179 |
| Rhouma *et al.* [14] | 7.9732 | 7.9750 | 7.9715 |
| Hongjun *et al.* [15] | 7.9851 | 7.9852 | 7.9832 |
| Huang *et al.* [16] | 7.8501 | 7.9028 | 7.5582 |
| Proposed | 7.9975 | 7.9942 | 7.9969 |

## 4. Conclusion

In this paper, a color image encryption algorithm based on chaotic maps is presented. The proposed algorithm is based on the concept of inter-component shuffling of image pixels using Arnold transform, then changing the gray values of the shuffled image using hyper-chaotic map. The control parameters of shuffling are extracted from the pending plain-image. The shuffling is inter-component scrambling i.e. the content of whole scrambled image doesn't change after scrambling, but it alters the contents of each components. The scrambled image is then encrypted using the preprocessed sequences generated by a 2D hyper-chaotic system. To get good encryption quality, the two key values are utilized to encrypt one pixel. The encryption is executed in CBC mode to make the process of confusion to encrypted image dependent. This way, the whole algorithm is made dependent to the pending plain-image, to make the potential cryptographic attacks infeasible, thereby improving the robustness of the algorithm against cryptanalysis. The experimental analysis and results demonstrate that the proposed algorithm has desirable properties like: high sensitivity to a small change in secret keys and plain-image, low correlation coefficients, low chi-square scores and large information entropy. All these features verify that the proposed algorithm is robust and effective for practical color image encryption.

**References**
[1] Lawande Q V, Ivan B R, Dhodapkar S D., 2005, "Chaos based Cryptography: A new approach to secure communications", *BARC Newsletter* 258.
[2] Pisarchik A N, Flores-Carmona N J, Carpio-Valadez M., 2006, "Encryption and decryption of images with chaotic map lattices", *CHAOS* American Institute of Physics 16, pp. 033118-033118-6.
[3] Mao Y, Chen G, Lian S., 2004, "A novel fast image encryption scheme based on 3D chaotic Baker maps", *International Journal of Bifurcation and Chaos* 14, pp. 3613–3624.

[4]   Xiang T, Liao X, Tang G, Chen Y, Wong K., 2006, "A novel block cryptosystem based on iterating a chaotic map", *Physics Letter A* 349, pp. 109-115.
[5]   Seyedzadeh S M, Mirzakuchaki S., 2012, "A fast color image encryption algorithm based on coupled two-dimensional piecewise chaotic map", *Signal Processing* 92, pp. 1202–1215
[6]   Zhong X, Liu J, Huang X., 2007, "An Image Encryption Algorithm Based on chaotic Cat Map", *Chinese Microelectronics & Computer* 24, pp. 131-134.
[7]   Franc-ois M, Grosges T, Barchiesi D, Erra R., 2012, "A new image encryption scheme based on a chaotic function", *Signal Processing: Image Communication* 27, pp. 249–259.
[8]   Akhshani A, Behnia S, Akhavan A, Hassan H A, Hassan Z., 2010, "A novel scheme for image encryption based on 2D piecewise chaotic maps", *Optics Communications* 283, pp. 3259–3266
[9]   Fu C, Lin B, Miao Y, Liu X, Chen J., 2011, "A novel chaos-based bit-level permutation scheme for digital image encryption", *Optics Communications* 284, pp. 5415–5423
[10]  Yao H, Li M., 2009, "An Approach of Image Hiding and Encryption Based on a New Hyper-chaotic System", *International Journal of Nonlinear Science* 7, pp. 379-384.
[11]  Wang X, Zhang W, Guo W, Zhang J., 2013, "Secure chaotic system with application to chaotic ciphers", *Information Sciences* 221, pp. 555–570
[12]  Wang X, Teng L, Qin X., 2012, "A novel colour image encryption algorithm based on chaos", *Signal Processing* 92, pp. 1101–1108
[13]  Kwok H S, Tang W K S., 2007, "A fast image encryption system based on chaotic maps with finite precision representation", *Chaos, Solitons and Fractals* 32, pp. 1518–1529.
[14]  Rhouma R, Meherzi S, Belghith S., 2009, "OCML-based color image encryption", *Chaos, Solitons and Fractals* 40, pp. 309-318.
[15]  Hongjun L, Xingyuan W., 2010, "Color image encryption based on one-time keys and robust chaotic maps", *Computer and Mathematics with Applications* 59, pp. 3320–7.
[16]  Huang C K, Nien H H., 2009, "Multi chaotic systems based pixel shuffle for image encryption", *Optics Communications* 282, pp. 2123–2127.
[17]  Liao T L, Tsai S H., 1987, "Adaptive synchronization of chaotic systems and its application to secure communication", *Chaos, Solitons and Fractals* 3, pp1387-1396.
[18]  Shannon C E., 1948, "A mathematical theory of communication", *Bell System Technical Journal* 27, 379–423 (part I), 623–856 (part II).